\documentclass[a4paper,11pt]{article}
\usepackage{pos}

\usepackage{wrapfig}

\title{Commissioning and Testing of IceAct Telescopes at the IceCube Neutrino Observatory}
\ShortTitle{IceAct: Commissioning and Testing}

\author{The IceCube Collaboration \\{\normalsize \normalfont(a complete list of authors can be found at the end of the proceedings)}\\}

\emailAdd{lars.heuermann@rwth-aachen.de}
\emailAdd{arunneelakandaiyer@hotmail.com}

\abstract{
IceAct is an array of imaging air Cherenkov telescopes located at the ice surface above the IceCube Neutrino Observatory. Each telescope features a silicon photomultiplier-based 61-pixel camera and a Fresnel-lens as imaging optic, resulting in a 12-degree field of view. The design is optimized to be operated in harsh environments, particularly at the South Pole. The setup will consist of seven telescopes in a so-called fly's eye configuration, increasing the field of view to 36°, and an additional telescope 200m apart for stereoscopic observations.
Rigorous testing procedures have been performed before deployment to ensure that operation under these conditions is possible, e.g. night sky observations and cold temperature tests. Furthermore, on-site calibrations are used to verify the accuracy and reliability of the installation.
We derive the geometric alignment of each IceAct telescope by comparing the directional reconstruction of muons measured with IceCube to the corresponding primary particle direction reconstruction from IceAct.
This contribution presents these testing procedures. Additionally, we present the on-site alignment calibration, including a Graph Neural Network reconstruction for the primary particle direction in IceAct, verification on Monte Carlo simulation, and the application to a commissioning dataset.

\vspace{4mm}

{\bfseries Corresponding authors:}

Arun Vaidyanathan$^{2*}$,
Lars Heuermann$^{1}$\\
{$^{1}$ \itshape RWTH Aachen University}\\
{$^{2}$ \itshape Marquette University}\\[4mm]
$^*$ Presenter
}

\ConferenceLogo{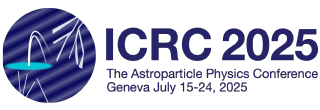}

\FullConference{39th International Cosmic Ray Conference (ICRC2025)\\
 15–24 July 2025\\
Geneva, Switzerland\\}

\begin{document}
\maketitle
\vspace{-0.2cm}
\section{Introduction}\label{sec1}
The next-generation of the IceCube experiment, known as IceCube-Gen2 \cite{gen2}, aims to advance multimessenger astronomy by exploring ultra-high energy neutrinos and other cosmic particles. This next-generation enhancement will allow us to probe the astrophysics of sources and fundamental physics at extreme-energy and distant sources \cite{tdr}. To achieve the ambitious goals, the detector systems are undergoing significant upgrades. These upgrades also foresee additional instrumentation, such as the Imaging Air Cherenkov Telescope Array known as IceAct.

\begin{wrapfigure}{r}{0.26\textwidth}
\includegraphics[width=1.0\linewidth]{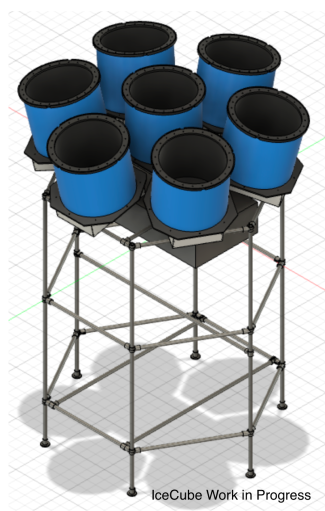}
\caption{CAD of an IceAct fly's eye telescope.}\label{fig:flyseye}
\end{wrapfigure}

Since 2019, two IceAct telescopes have taken data \cite{Andeen:2019AD} at the IceCube Neutrino Observatory \cite{Aartsen:2016nxy}.
The telescopes feature a SiPM-based 61-pixel camera and a Fresnel lens, resulting in a 12-degree field of view. For more details on the hardware implementation, see \cite{threeyear}. The IceAct telescopes help identify the mass composition over a wide energy range (about 100~TeV to 10~PeV) by reconstructing extensive air shower (EAS) parameters. Another use-case is the cross-calibration of energy and directional measurements of IceTop and the IceCube in-ice detectors. Since 2025, five IceAct telescopes have been installed at IceCube. One is mounted on the rooftop of the IceCube Laboratory (referred to as \emph{roof}), while the others are positioned on the ice surface, 220~meters west-southwest, near the initial position of the 2019 telescope (referred to as \emph{field}). The field telescopes are being arranged into a fly's eye configuration (see \autoref{fig:flyseye}), in which the remaining three telescopes are presently built and being tested, and are expected to be deployed in an upcoming season to complete the first fly's eye upgrade.
\vspace{-0.3cm}

\section{Telescope Test Procedures}

The testing of the telescope involves a series of Laboratory and on-ice test procedures designed to verify its performance, reliability, and readiness for deployment under harsh environmental conditions. Laboratory tests are conducted during the assembly phase, while the final validation of the telescope alignment is performed using on-ice observation and simulation data.

\subsection{Laboratory Test Procedure}

The Laboratory tests includes five key steps: capacitance tests, trigger scanning, performance tests, freezer tests, and night sky observations.

\begin{figure}[t!]
\includegraphics[width=0.5\linewidth]{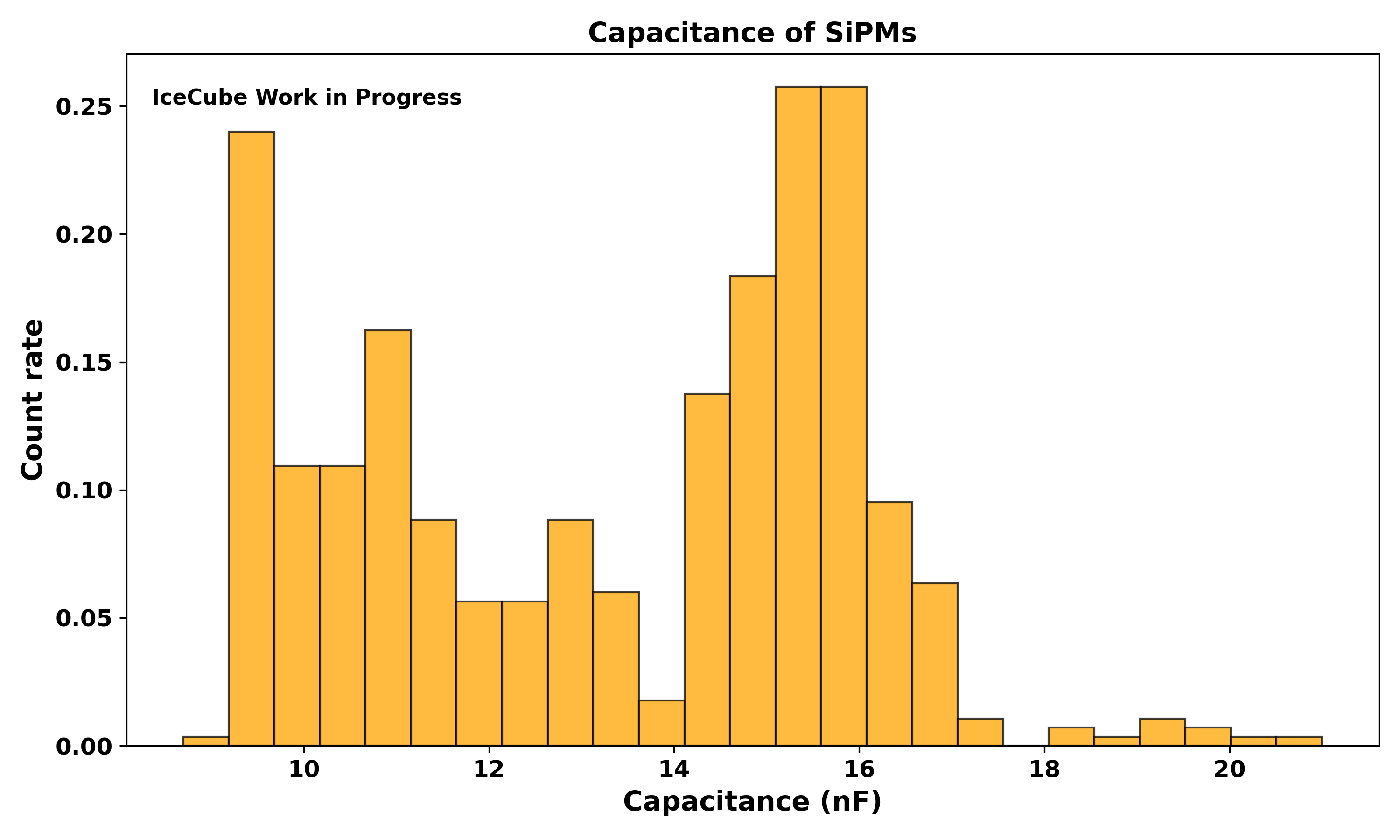}
\includegraphics[width=0.545\linewidth]{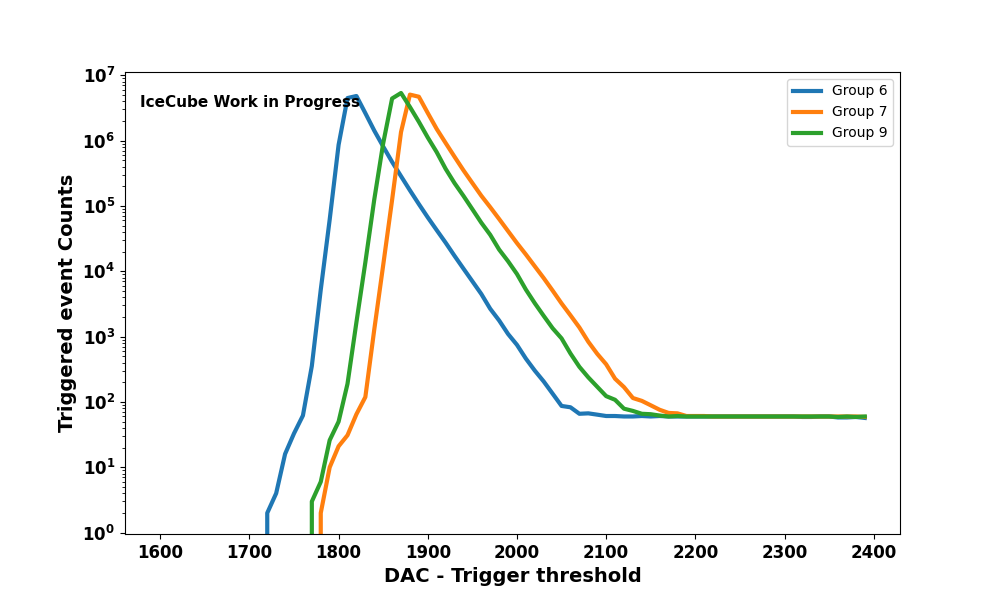}
\caption{Histogram of measured capacitance values of nine new camera boards (left). Triggering behavior (right).}\label{fig01}
\end{figure}

\textbf{Capacitance Tests:} Capacitance testing is a critical diagnostic step for identifying malfunctioning or incorrectly mounted SiPMs (Silicon Photo-Multipliers). Properly functioning SiPMs typically exhibit capacitance values between 5~nF and 30~nF (\autoref{fig01}). If an SiPM is rotated 180$\textdegree$ during soldering, it may not be visually identifiable, but can be easily detected via this test. This method is also useful for identifying damaged SiPMs.

\textbf{Trigger scanning:} This test is performed using an external light source to evaluate the response characteristics of the SiPMs and the associated electronics on the camera board. In this procedure, the number of events triggered is recorded over a fixed time window, typically 0.2~s, by setting a constant reference voltage \cite{threeyear}. The outcome of this test is a characteristic trigger efficiency curve for each trigger group (\autoref{fig01}). The width of the curve indicates the level of electronic noise. Irregularities or abnormal shifts in the curve may indicate problems with the preamplification circuit or issues with the SiPMs. Comparison of curves across multiple trigger groups helps to isolate the faults in individual channels.

\begin{figure}[t!]
\centering
\includegraphics[width=0.8\linewidth]{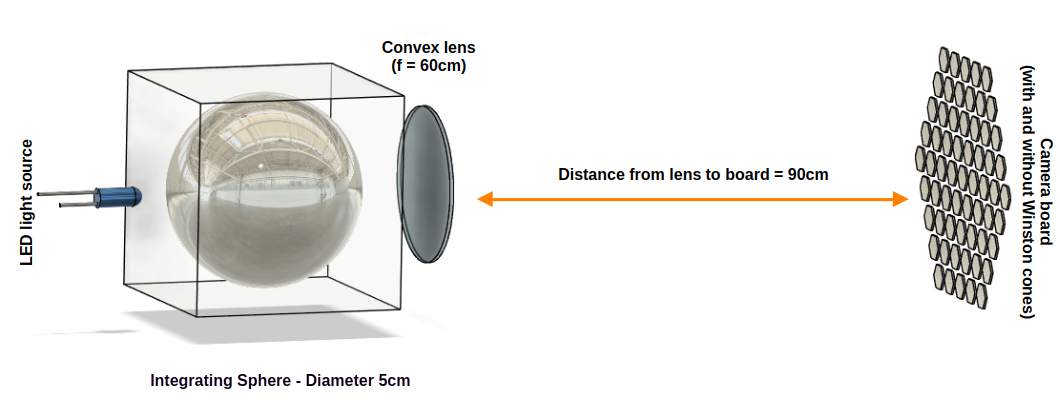}
\caption{Schematic representation of performance test setup arrangement.}\label{fig02}
\end{figure}

\begin{figure}[t!]
\centering
\includegraphics[width=0.95\linewidth]{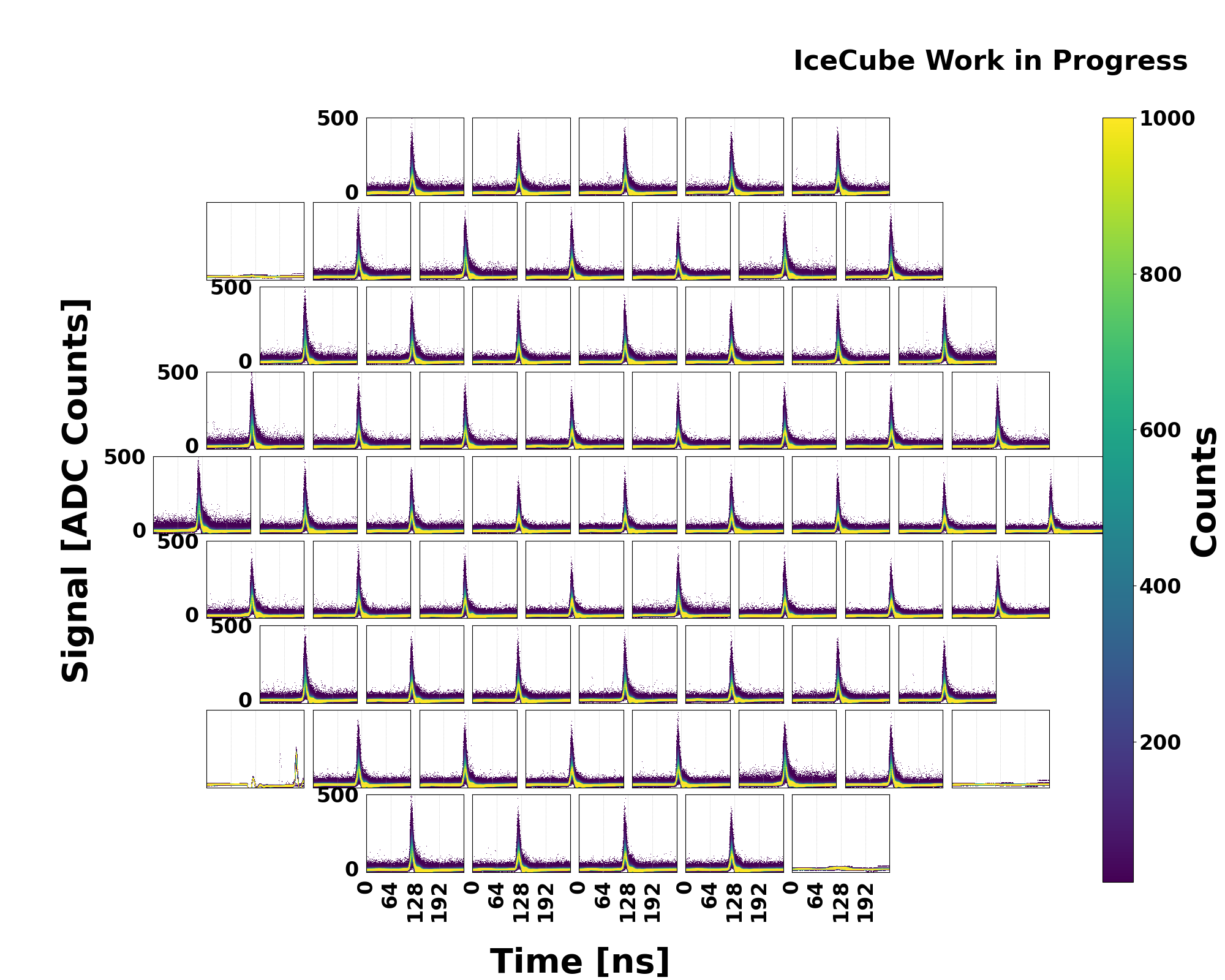}
\caption{Resulting spectra of darkroom calibration test using a new camera board (without Winston cones).}\label{fig04}
\end{figure}

\textbf{Performance Tests:} Performance verification is carried out in a darkroom using a calibrated light source coupled to an integrating sphere, which ensures uniform light diffusion across the camera surface. A convex lens is positioned in front of the integrating sphere to optimize light distribution (\autoref{fig02}). The distance between the camera board and the light source is maintained constant for all testing procedures to ensure consistency and comparability of the results. The acquired data is calibrated and analyzed to generate pixel spectra (\autoref{fig04}), enabling the identification of inconsistencies or defective pixels or preamplifier components on the camera board.

\begin{figure}[t!]
\centering
\includegraphics[width=0.8\linewidth]{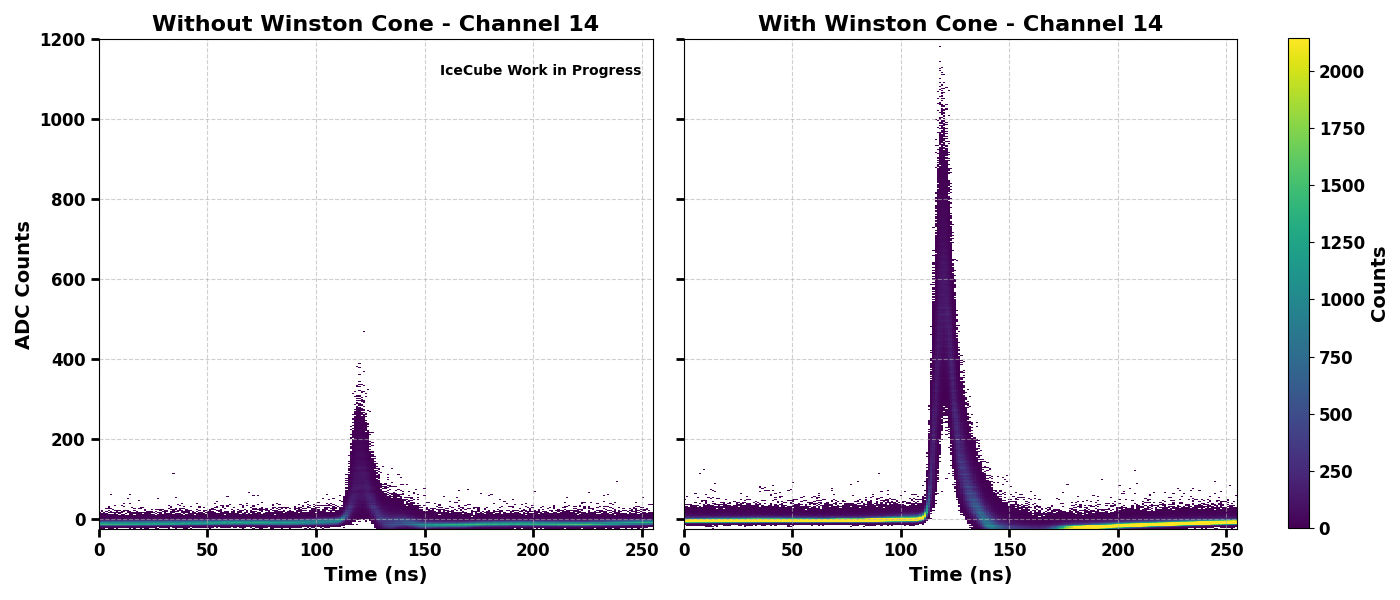}
\caption{Comparison of spectrum of performance test before and after attaching Winston cones.}\label{fig05}
\end{figure}

Following this, Winston cones are attached to each pixel to enhance light collection efficiency. The performance test is then repeated to verify the expected improvement in photon collection due to the cones (\autoref{fig05}). To avoid the challenges of reworking the camera board after the Winston cones are attached, it is essential to perform testing both before and after cone installation.

\textbf{Freezer Tests:} To replicate the extreme temperatures at the South Pole, the camera board is subjected to a cold endurance test. The camera board was placed in a freezer at -60\textdegree~C for about 24~hours, then brought back to room temperature. This cycle is repeated two to three times to confirm the stability of the Winston cone attachments and to ensure that there is no loosening or damage due to thermal stress. Additional low-temperature performance tests are conducted after assembling the full telescope to confirm proper functionality of the entire system under South Pole-like conditions.

\textbf{Night Sky Observations:} The final phase of testing involves observation under a clear night sky. At least two hours of data are collected from a rooftop arrangement at the physics buildings of RWTH Aachen University and Marquette University. During the night sky test, ambient light from streetlights was minimized by surrounding the telescope with a barrier made of hardboard or tarp walls. This shielding significantly reduced the background noise level, ensuring cleaner data collection. This step validates the performance in a realistic observational environment and confirms that the system is ready for deployment.
\vspace{-0.2cm}
\section{On-Ice Test Procedure}

\begin{wrapfigure}{r}{0.26\textwidth}
    \centering
    \includegraphics[width=0.25\textwidth]{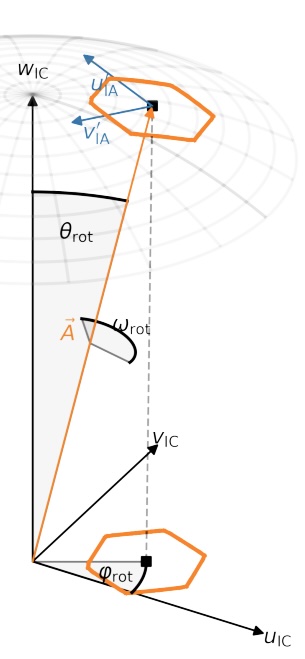}
    \caption{Visualisation of the orientation as Euler angles, together with the field of view of an IceAct telescope (orange). The pointing vector of the telescope is denoted as $\vec{A}$.}
    \label{fig:alignment}
\end{wrapfigure}

To verify the performance of the IceAct telescopes at the South Pole, we derive the relative orientation of the telescopes to the IceCube coordinate system.
For example, the IceAct fly's eye configuration results in the frame of reference of a singular telescope being tilted and rotated against the IceCube frame of reference.
If we express the direction reconstruction of the zenith $\theta$ and the azimuth $\phi$ as cartesian vector $(u,v,w)^T = (\sin(\theta)\cos(\phi),\sin(\theta)\sin(\phi),\cos(\theta))^T$, the transformation into the IceCube Coordinate system can be expressed as $\mathbf{u}= \mathbf{M}\cdot\mathbf{u'}$, where $\mathbf{M}$ is a $3\times3$ rotation matrix, here referred to as orientation.

The orientation can be intuitively described by Euler angles, as depicted in \autoref{fig:alignment}. The angle
$\theta_{\mathrm{rot}}$ is the inclination along the zenith, $\phi_{\mathrm{rot}}$ is the direction angle of this inclination along the azimuth, and $\omega_\mathrm{rot}$ expresses the rotation angle of the telescope along its pointing direction.
Accurate determination of these angles (sub-degree precision) is crucial to minimize systematic uncertainties in the reconstructions, which are important for cross-calibration between IceCube and IceAct.
This is achieved with a common hybrid data sample.
We select IceCube in-ice and IceAct telescope data based on timing and trigger information.
For the measured in-ice muon tracks of these events, we reconstruct the direction of the muon. This direction is closely related to the arrival direction of the primary cosmic-ray particle. In a second step, we compare the reconstructed muon direction to the shower arrival direction reconstructed with IceAct and fit the orientation.

The comparison of reconstructed directions in IceAct and IceCube is done on an event-by-event basis with a likelihood fit.
The likelihood itself is based on the von-Mises-Fisher distribution \cite{vMF}. The resulting log-likelihood (up to a constant) is:
\begin{equation}
    \label{eq:likelihood}
    -2\ln{\mathcal{L}}\left(\mathbf{x}|q\right) = -2 \sum^\mathrm{n_{events}} \kappa \mathbf{u_{IC}}^T\mathbf{M}(q) \mathbf{u'_{IA}},
\end{equation}
where $\mathbf{u_{IC}}$ is the cartesian unit vector of the in-ice muon direction, $\mathbf{u'_{IA}}$ is the corresponding IceAct reconstruction in the reference system of IceAct, $\kappa$ is the so-called concentration parameter of the von-Mises-Fisher distribution, and $\mathbf{M}(q)$ is the orientation, expressed in dependence of the unit-quaternion $\mathbf{q} = \left(x_0\sin{\vartheta}, x_1\sin{\vartheta},x_2\sin{\vartheta}, \cos{\vartheta}\right)$. The vector $\mathbf{x}=(x_0,x_1,x_2)$ thereby denotes the axis on which a telescope is rotated around by the amount of $\vartheta$. Expressing the rotation as a quaternion increases the numerical stability and avoids the gimbal lock of Euler angles.

\subsection{Evaluation on Monte Carlo Data}

\begin{wrapfigure}{r}{0.35\textwidth}
    \centering
    \includegraphics[width=0.3\textwidth]{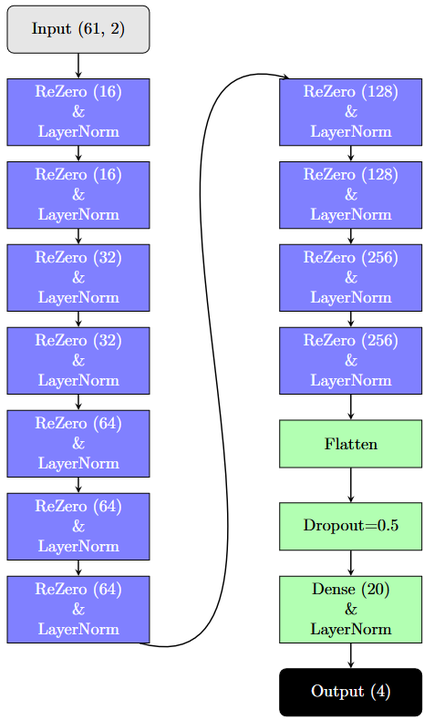}
    \caption{GCNN Architecture used for the directionality reconstruction}
    \label{fig:GCNN}
\end{wrapfigure}

A joint Monte Carlo simulation \cite{ICRC2023Larissa} of the telescopes and the IceCube detector is used in two ways: to evaluate the performance of the fit, and to train a graph-convolutional neural network (GCNN) to reconstruct the track direction in the IceAct field of view.
We simulate the position of eight different fly's eye configurations with seven telescopes each; more details can be found in \cite{ICRC2023Larissa, IceAct_Spectrum}.

\begin{figure}[b]
    \centering
    \includegraphics[width=0.49\linewidth]{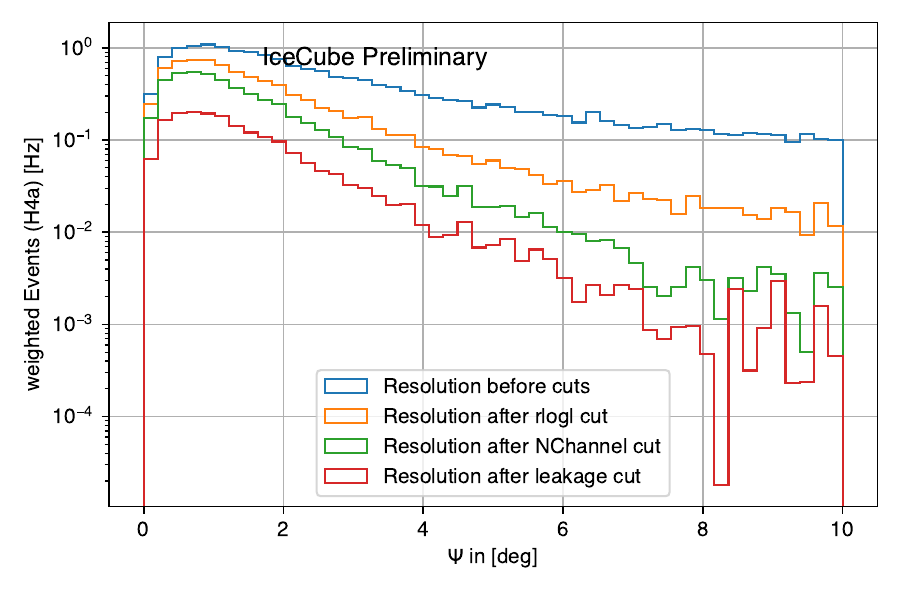}
    \includegraphics[width=0.49\linewidth]{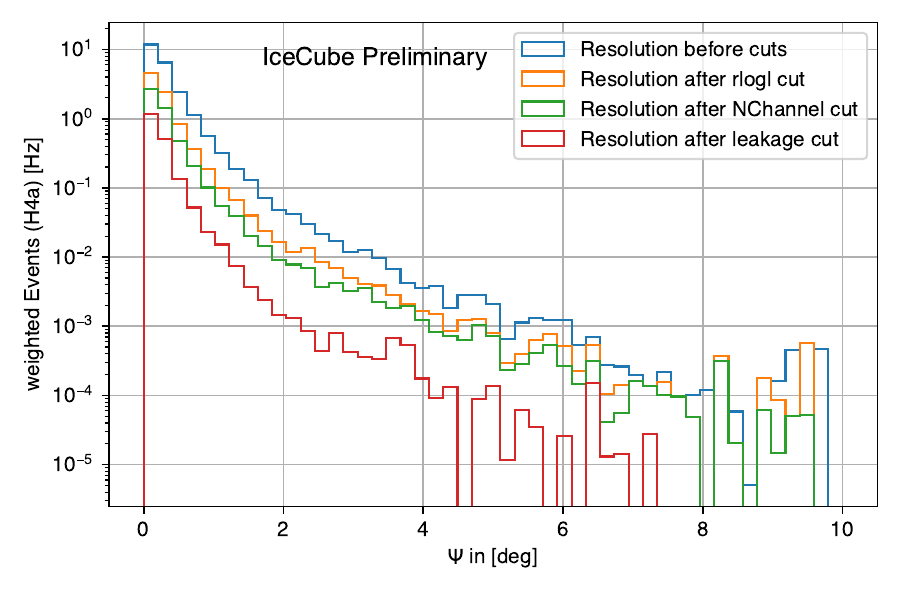}
    \caption{\textbf{Left}: Histogram of the opening angle $\Psi$ between SplineMPE reconstruction and true simulated primary particle direction after the application of quality cuts.\\
    \textbf{Right}: Histogram of the opening angle $\Psi$ between the GCNN reconstruction and the primary particle direction for the same quality cuts.}
    \label{fig:QualityCutsSplineMPE}
\end{figure}

We use SplineMPE \cite{SplineMPE} (a standard IceCube algorithm) for the in-ice muon reconstruction.
We apply a number of basic quality cuts:
the number of hit DOMs (NChannel) has to be above 50, and the likelihood per degree of freedom (rlogl) has to be below 7.5.
Additionally, we exclude events that are only partially contained within the IceAct camera: if the fraction of the summed amplitude measured in the outermost camera-pixel-ring (leakage) is above 0.05, we reject the event. The figure \autoref{fig:QualityCutsSplineMPE} shows that the cuts reduce the opening angle between truth and reconstruction for SplineMPE as well as the neural network.

As a final quality cut, we propagate the reconstructed muon track back to the ice-surface and apply a 300-meter-radius cut on a telescope-by-telescope basis around the median position of the tracks.

The GCNN architecture for reconstructing the IceAct direction is based on \cite{ICRC2023Larissa}. The architecture has been optimized for application in directional reconstruction. The network heavily utilizes ReZero \cite{ReZero} residual layers, which allow a deeper neural network architecture.
Furthermore, we use layer normalisation. This procedure numerically stabilizes the latent space of the neural network and opens the possibility for feature extraction for a future multi-telescope neural network reconstruction, as used in the Advanced Northern Track Selection~\cite{ANTS_ICRC}. 
A von-Mises-Fisher distribution is used as the loss function. The GCNN predicts normalized direction vectors as well as $\kappa$ for each event, resulting in 4 output parameters. 
The complete architecture is shown in \autoref{fig:GCNN}. The resulting median opening angle between Monte-Carlo truth and reconstruction is below 0.2 degrees (see \autoref{fig:QualityCutsSplineMPE}), well below the single pixel FoV of 1.5\textdegree and below the median resolution of SplineMPE.

\begin{figure}
    \centering
    \includegraphics[width=0.55\linewidth]{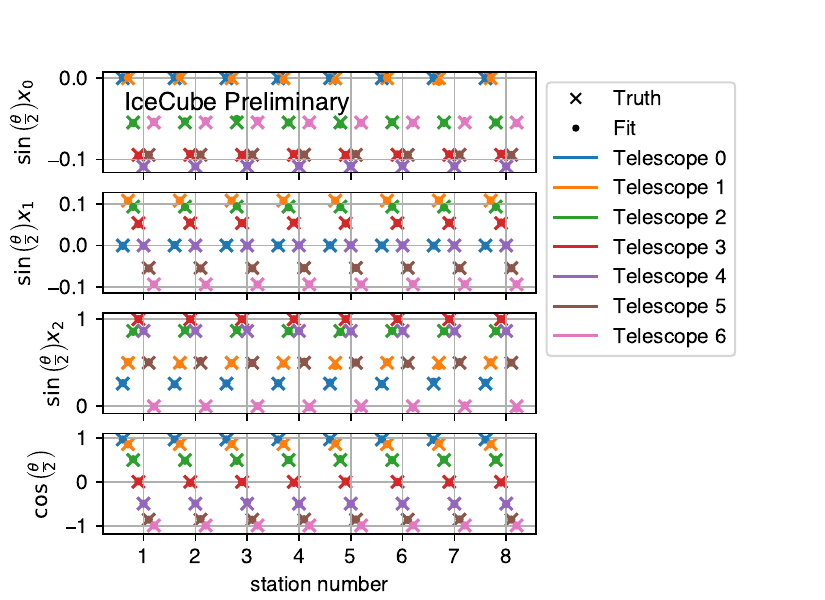}
\includegraphics[width=0.4\textwidth]{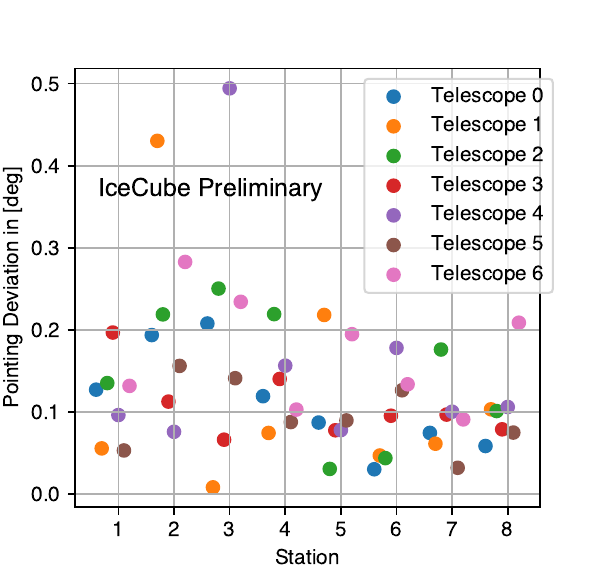}
\caption{\textbf{Left:} Results of the orientation fit for each telescope, as well as the simulated Monte Carlo truth. \textbf{Right:} Opening angle between MC truth and reconstructed pointing direction for each telescope.}
        \label{fig:FitResults}
\end{figure}

We perform a likelihood fit for all seven telescopes of all eight simulated fly's eye configurations. The result can be seen in \autoref{fig:FitResults}. The true quaternion is recovered for each telescope and each fly's eye configuration. The precision achieved is well below one degree, with an average below 0.3\textdegree. Especially for the case of the zenith-pointing telescope "0", the worst deviation is 0.2\textdegree, thus the method can be applied to commission data of the two zenith-pointing telescope of 2021.

\vspace{-0.2cm}

\subsection{Application on Commission Data}

\begin{figure}
  \begin{minipage}[c]{0.7\textwidth}
    \includegraphics[width=\textwidth]{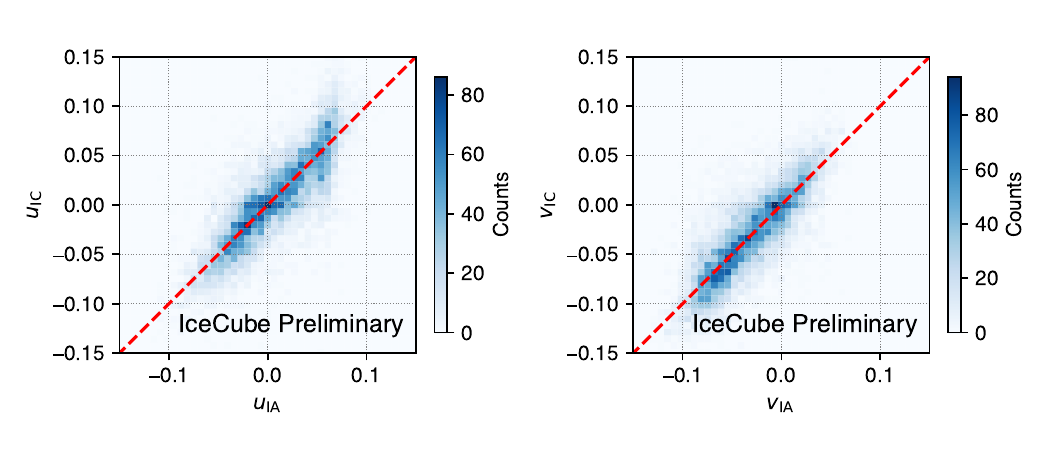}
  \end{minipage}\hfill
  \begin{minipage}[c]{0.3\textwidth}
    \caption{
     Comparison between the IceCube and IceAct reconstruction. The quantities $u_{IC}$ and $v_{IC}$ are the reconstructed coordinates of IceCube. Similarly, $u_{IA}$ and $v_{IA}$ refer to the IceAct reconstruction rotated in the coordinate system of IceCube.
    } \label{fig:SplineMPEvsIceAct}
  \end{minipage}
\end{figure}

The final likelihood fit is performed for the roof and field telescopes for a week of commissioning data in 2021.
The resulting Euler angles for the roof and field are: ~mbox{$
\theta_\mathrm{field} = 0.3^\circ,\
\phi_\mathrm{field}= 32.3^\circ,\
\omega_\mathrm{field}=71.4^\circ
$}
and
\mbox{$
\theta_\mathrm{roof} = 1.7^\circ, \
\phi_\mathrm{roof}= -92.9^\circ, \
\omega_\mathrm{roof}=-17.1^\circ.$} With this derived orientation, we can now rotate the prediction of IceAct in the frame of reference of IceCube and compare the in-ice reconstruction the IceAct reconstruction, this can be seen in \autoref{fig:SplineMPEvsIceAct}, for both of the Cartesian coordinate vectors.

The tilt of the telescope is easily visible, as the distribution of the $v$ coordinnate is not centered around 0. In general, the predictions of SplineMPE and the GNN agree well, and the statistical uncertainties of the reconstruction methods dominate the uncertainty on the estimated direction.
\vspace{-0.3cm}
\section{Conclusion and Outlook}\label{sec3}
\vspace{-0.05cm}
In this proceeding contribution, we presented the design, testing, and commissioning procedures of the IceAct telescopes at the IceCube Neutrino Observatory, which were deployed in 2023 and 2024. Laboratory tests confirmed the reliability of the SiPM camera boards, electronics, and mechanical structures under extreme conditions. The orientation fit for two telescopes with a week-long dataset from 2021 demonstrated that with the method is now ready to be applied to the full dataset of a complete fly's eye configuration. A comprehensive analysis of the systematic uncertainties of the orientation, including time-dependent variations, is currently in preparation.
The application of GCNN to other variables, such as energy, shower core position, and shower maximum, as well as the inclusion of multiple telescopes and detector reconstruction, is ongoing work. This ongoing work will refine IceAct’s reconstruction capabilities and integration within IceCube.

\vspace{-0.2cm}
\bibliographystyle{ICRC}
\bibliography{skeleton.bbl}

\clearpage

\section*{Full Author List: IceCube Collaboration}

\scriptsize
\noindent
R. Abbasi$^{16}$,
M. Ackermann$^{63}$,
J. Adams$^{17}$,
S. K. Agarwalla$^{39,\: {\rm a}}$,
J. A. Aguilar$^{10}$,
M. Ahlers$^{21}$,
J.M. Alameddine$^{22}$,
S. Ali$^{35}$,
N. M. Amin$^{43}$,
K. Andeen$^{41}$
% ...
\\
\\
% ...
$^{10}$ Universit{\'e} Libre de Bruxelles, Science Faculty CP230, B-1050 Brussels, Belgium \\
% ...
$^{16}$ Department of Physics, Loyola University Chicago, Chicago, IL 60660, USA \\
$^{17}$ Dept. of Physics and Astronomy, University of Canterbury, Private Bag 4800, Christchurch, New Zealand \\
% ...
$^{21}$ Niels Bohr Institute, University of Copenhagen, DK-2100 Copenhagen, Denmark \\
$^{22}$ Dept. of Physics, TU Dortmund University, D-44221 Dortmund, Germany \\
% ...
$^{35}$ Dept. of Physics and Astronomy, University of Kansas, Lawrence, KS 66045, USA \\
% ...
$^{39}$ Dept. of Physics and Wisconsin IceCube Particle Astrophysics Center, University of Wisconsin{\textemdash}Madison, Madison, WI 53706, USA \\
% ...
$^{41}$ Department of Physics, Marquette University, Milwaukee, WI 53201, USA \\
% ...
$^{43}$ Bartol Research Institute and Dept. of Physics and Astronomy, University of Delaware, Newark, DE 19716, USA \\
% ...
$^{63}$ Deutsches Elektronen-Synchrotron DESY, Platanenallee 6, D-15738 Zeuthen, Germany \\
$^{\rm a}$ also at Institute of Physics, Sachivalaya Marg, Sainik School Post, Bhubaneswar 751005, India \\
% ...

\subsection*{Acknowledgments}

\noindent
The authors gratefully acknowledge the support from the following agencies and institutions:
USA {\textendash} U.S. National Science Foundation-Office of Polar Programs,
U.S. National Science Foundation-Physics Division,
U.S. National Science Foundation-EPSCoR,
U.S. National Science Foundation-Office of Advanced Cyberinfrastructure,
Wisconsin Alumni Research Foundation,
Center for High Throughput Computing (CHTC) at the University of Wisconsin{\textendash}Madison,
Open Science Grid (OSG),
Partnership to Advance Throughput Computing (PATh),
Advanced Cyberinfrastructure Coordination Ecosystem: Services {\&} Support (ACCESS),
Frontera and Ranch computing project at the Texas Advanced Computing Center,
U.S. Department of Energy-National Energy Research Scientific Computing Center,
Particle astrophysics research computing center at the University of Maryland,
Institute for Cyber-Enabled Research at Michigan State University,
Astroparticle physics computational facility at Marquette University,
NVIDIA Corporation,
and Google Cloud Platform;
Belgium {\textendash} Funds for Scientific Research (FRS-FNRS and FWO),
FWO Odysseus and Big Science programmes,
and Belgian Federal Science Policy Office (Belspo);
Germany {\textendash} Bundesministerium f{\"u}r Forschung, Technologie und Raumfahrt (BMFTR),
Deutsche Forschungsgemeinschaft (DFG),
Helmholtz Alliance for Astroparticle Physics (HAP),
Initiative and Networking Fund of the Helmholtz Association,
Deutsches Elektronen Synchrotron (DESY),
and High Performance Computing cluster of the RWTH Aachen;
Sweden {\textendash} Swedish Research Council,
Swedish Polar Research Secretariat,
Swedish National Infrastructure for Computing (SNIC),
and Knut and Alice Wallenberg Foundation;
European Union {\textendash} EGI Advanced Computing for research;
Australia {\textendash} Australian Research Council;
Canada {\textendash} Natural Sciences and Engineering Research Council of Canada,
Calcul Qu{\'e}bec, Compute Ontario, Canada Foundation for Innovation, WestGrid, and Digital Research Alliance of Canada;
Denmark {\textendash} Villum Fonden, Carlsberg Foundation, and European Commission;
New Zealand {\textendash} Marsden Fund;
Japan {\textendash} Japan Society for Promotion of Science (JSPS)
and Institute for Global Prominent Research (IGPR) of Chiba University;
Korea {\textendash} National Research Foundation of Korea (NRF);
Switzerland {\textendash} Swiss National Science Foundation (SNSF). The Wisconsin Space Grant Consortium and the Northwestern Mutual Data Science Institute.

\end{document}